# Electrocaloric response in Lanthanum-modified lead zirconate titanate ceramics

B. Asbani[1,2], M. El Marssi[1], J.-L. Dellis[1], A. Lahmar[1], Y. Gagou[1], D. Mezzane[2], M. Amjoud[2], A. Alimoussa[2], Z. Kutnjak[3], R. Pirc[3], and B. Rožič[3,*]

[1]*LPMC, Université de Picardie Jules Verne, 33 rue Saint-Leu, 80039 Amiens Cédex, France*

[2]*LMCN, F.S.T.G. Université Cadi Ayyad, BP 549, Marrakech, Morocco*

[3]*Jozef Stefan Institute, Jamova cesta 39, 1000 Ljubljana, Slovenia*

**Abstract**

Recent findings of a large electrocaloric (EC) effect in polymeric and inorganic ferroelectric materials open a potential possibility of development of solid-state cooling or heating devices of new generation with better energy efficiency that may be less harmful for the environment. We investigate by using direct measurements, the temperature and electric field dependence of the electrocaloric response in $Pb_{1-x}La_x(Zr_yTi_{1-y})_{1-x/4}O_3$ bulk ceramics (PLZT) with $x$=0.06 and 0.12. Here, the properties of the EC response were probed in a part of the PLZT composition phase diagram with low $y$=0.40 composition, in which the EC effect was not previously studied. Measurement results show the existence of the sizeable EC response in 12/40/60 PLZT sample with the EC temperature change ($\Delta T_{EC}$) of 2.92 K at 430 K and 80 kV/cm. This value exceeds previously obtained $\Delta T_{EC}$ values in relaxor ferroelectric x/65/35 PLZT compositions and rivaling the best EC response in lead magnesium niobate-lead titanate ceramics. The electrocaloric responsivity ($\Delta T/\Delta E$) value of 0.41x$10^{-6}$ Km/V determined at a lower electric field of 20 kV/cm and 410 K is comparable to those observed in other perovskite ferroelectrics.

* brigita.rozic@ijs.si

**I. INTRODUCTION**

Relatively small electrocaloric effect (ECE) observed in ferroelectrics over the past decades hindered the development of new dielectric cooling technology based on ECE. However, the recent discovery of a giant electrocaloric effect in some organic and inorganic ferroelectric materials [1-10] provided significant momentum in the revival of the research of the ECE [1-4,11-20]. Besides, prototypes of electrocaloric cooling devices and theoretical work related to the efficiency of such devices demonstrate the potential possibility to develop solid-state heat-management devices with high energy efficiency [17,21-25]. The mechanism underlying the electrocaloric (EC) effect revolves around the change of entropy stimulated by the electric field induced change of the ferroelectric dipolar state [2,26-28].

Indirect electrocaloric experiments demonstrated that the giant electrocaloric effect exists in both inorganic and organic ferroelectric materials, i.e., in a thin $PbZr_{0.95}Ti_{0.05}O_3$ (PZT) [2] film and thick poly(vinylidene fluoride-trifluoroethylene) (P(VDF-TrFE)) copolymer films [3], respectively. Later on, indirect and direct measurements performed on bulk samples demonstrated that significant electrocaloric effect exists in various perovskite ferroelectric relaxor materials [1,9-12,17], which are very interesting for application due to their large dielectric and electromechanical properties [29]. Besides, large ECE was also demonstrated in thin and thick perovskite ceramic films [1,8,9,12,30].

It is interesting to note that both positive and negative large ECE can be found in ferroelectric (FE) and antiferroelectric (AFE) materials [28-35]. Recently, it was shown that the enhanced electrocaloric effect could be found in perovskite relaxor ferroelectrics in the vicinity of the critical point of the liquid-vapor type [30]. This include the relaxor compositions of $Pb_{1-x}La_x(Zr_yTi_{1-y})_{1-x/4}O_3$ ceramics with $0.05<x<0.12$ and $y=0.65$ (x/65/35 PLZT ceramics) [20,32]. In particular, for $[PbMg_{1/3}Nb_{2/3}O_3]_{0.90}$-$[PbTiO_3]_{0.10}$ (PMN-10PT) ceramics the electrocaloric temperature change, $\Delta T_{EC}$ of 3.45 K was reported [36], while in 8/65/35 PLZT, the ECE of 2.25 K was observed [30]. Such large electrocaloric values can already be exploited in electrocaloric cooling devices [17,21].

While several compounds of La-modified compositions x/65/35 PLZT ceramics were investigated, other compositions attracted little or no attention, except for antiferroelectric compositions 2-5/95/5 PLZT [37-39], 11/70/30 and 7/82/18 PLZT ceramics [40]. In thin films of 2/95/5 PLZT ceramics, the ECE exceeding 6 K was found at relatively large electric fields of 300 kV/cm [38]. Furthermore, in bulk 11/70/30 and 7/82/18 PLZT ceramics, the ECE was not exceeding the 1.5 K for fields below 70 kV/cm. However, theoretical calculations indicate that large ECE exceeding 7 K could be potentially found in these compositions [40].

In this work, we investigate by direct electrocaloric measurements the temperature and electric field dependence of the electrocaloric response in an entirely different part of the PLZT composition phase diagram with low $y$=0.40 composition. Here, the bulk ceramics with $x$=0.06 and 0.12 (denoted by 6/40/60 and 12/40/60 PLZT ceramics) were chosen in which ECE was not studied previously. The electrocaloric effect was investigated in the temperature range from 310 K up to 430 K, i.e., near the room temperature. The results demonstrate the importance of the proximity and sharpness of the paraelectric to ferroelectric phase transition for the electrocaloric response enhancement.

As demonstrated in Refs. [26,27] the electrocaloric temperature change ($\Delta T_{EC}$) is obtained from the equation

$$\frac{\Delta T_{EC}}{T} = exp\left\{\frac{P^2(0,T) - P^2(E, T+\Delta T_{EC})}{2C_{ph}(T)}\right\} \quad (1)$$

as a function of the electric field ($E$) and temperature ($T$) for ferroelectrics or relaxor ferroelectrics. Here, the $C_{ph}$ represents the lattice heat capacity and the polarization ($P$) is calculated from the equations of state based on the standard form of the free energy ($f$) expressed as [26]

$$f = f_0 + \frac{1}{2}aP^2 + \frac{1}{4}bP^4 + \frac{1}{6}cP^6 - PE, \quad (2)$$

where the Landau-Ginzburg-Devonshire type expansion coefficients are denoted by $a = a_1(T - T_0)$, $b$, and $c$, with $T_0$ the paraelectric-to-ferroelectric transition temperature in zero field. The extensive calculations for relaxor ferroelectrics and ferroelectrics have shown that the maximum ECE response is expected at the paraelectric to ferroelectric phase transition [26,27]. In some relaxor ferroelectrics, such transitions are only induced by the strong enough electric field [29,41]. Here we will investigate the 6/40/60 PLZT ceramics with the paraelectric to ferroelectric (P-FE) phase transition at 570 K, i.e., the temperature well above of the investigated temperature range and 12/40/60 PLZT ceramics with the P-FE phase transition at 390 K, i.e., the temperature within the investigated temperature range.

## II. EXPERIMENTAL

The ECE was determined by direct electrocaloric measurements in bulk 6/40/60 and 12/40/60 PLZT ceramics that were prepared by a conventional mixed-oxide method followed by a hot-press sintering at 1250 °C and a final half-an-hour annealing at 600 °C to release all internal mechanical stresses [42-45]. All prepared samples were single-phase perovskites determined by the X-ray diffraction analysis [43]. These materials are known to possess interesting dielectric and pyroelectric properties [43]. Additional characterization data of these PLZT ceramics can be found in Refs. [42-45].

The resulting PLZT platelets were cut and thinned down to a thickness of 64-130 μm, producing pairs with similar thickness for both compositions. Gold electrodes covered both surfaces (see inset to Fig. 1). The details of the experimental method and data analysis can be found in Refs. [8,9,30]. Before each measurement, the samples were annealed at 300 °C to release pinning oxygen vacancies and to anneal further any mechanical stresses. Here, the direct EC measurement method was based on a modified high-resolution calorimeter allowing precise temperature stabilization of the bath (within ± 0.1 mK). Besides, it enables high-resolution measurements of the sample temperature variation related to the electrocaloric effect, which was induced by the change of the applied electric field [30]. Temperature was measured by using a small bead thermistor glued to one sample's electrode (see inset to Fig. 1).

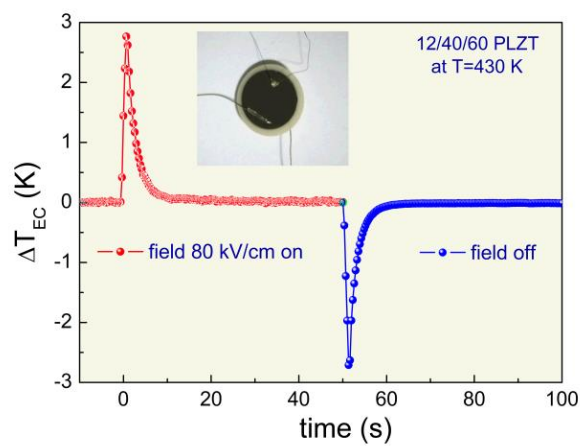

FIG. 1. The measured ECE signal as a function of time in bulk 12/40/60 PLZT ceramics. The inset shows a typical sample arrangement, including the wires attachment and the small bead thermistor.

The electric field was applied in the form of step-like pulses, always starting from zero. The length of the pulses was exceeding both the internal and external thermal equilibrium time scales of 0.5-1 s and 5-10 s, respectively. Short internal thermal response time allowed the application of the simple zero-dimensional model with sufficient accuracy [8,9,30]. From the relaxation of the temperature of the completely internally equilibrated system back to the bath temperature (see data in Fig. 1), the amplitude $\Delta T$ of the measured ECE is determined from fitting $T(t)$ to the ansatz $T(t) = T_{bath} + \Delta T e^{-t/\tau}$. Since only the part of the sample covered with electrodes is exhibiting the ECE, the actual heat change of the ECE participating part was determined from $\Delta T_{EC} = \Delta T \sum_i C_p^i / C_p^{EC}$ [8,9,30]. Here, the geometry of the system and the heat capacities of its constituents $C_p^i$ (the thermistor, attaching wires, glue, gold electrodes, etc.) were taken into account. The $C_p^{EC}$ denotes the heat capacity of the EC active part of the sample, i.e., electrode part. The total heat capacity of the attached materials to the sample was below 5%. Taking into account all errors in the determination of $C_p^i$ and $\Delta T$, the total error of $\Delta T_{EC}$ is estimated to be below 3%. Symmetry of data peaks in Fig. 1 and systematic return of ECE data back to zero at longer times demonstrate the absence of the Joule heating in x/40/60 PLZT ceramics with x=0.06 and 0.12.

Note that due to the relatively thick 12/40/60 PLZT sample, only one measurement at 80 kV/cm was conducted to preserve the delicate electrical wiring of the calorimeter.

## III. RESULTS AND DISCUSSION

### A. Dielectric investigations

Fig. 2 shows the real part of the complex dielectric constant obtained at 10 kHz as a function of temperature for both PLZT compositions. The magnitude of the dielectric constant deep within the FE phase is within the expected small values for 6/40/60 PLZT sample, i.e., well below the transition temperature close to 570 K. No significant hysteresis was observed between cooling and heating runs that far from the transition.

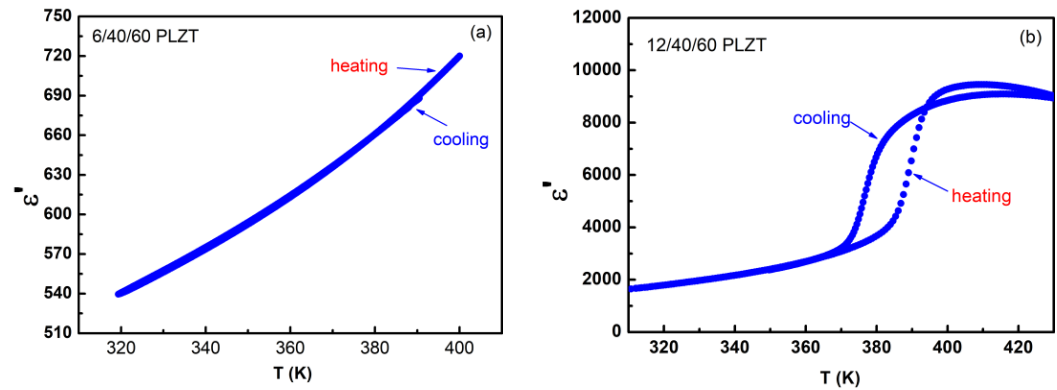

FIG. 2. Temperature dependence of the real part of the dielectric constant ε' measured at 10 kHz in bulk 6/40/60 PLZT ceramics (a) and bulk 12/40/60 PLZT ceramics (b).

In contrast, the dielectric temperature dependence clearly shows the FE phase transition near 390 K for the 12/40/60 PLZT sample. Visible hysteresis between cooling and heating runs indicate a first-order nature of the FE transition. It is expected that latent heat released or absorbed at this transition can significantly enhance the ECE response [1,10,32].

### B. Electrocaloric investigations

The temperature dependence of the ECE temperature change $\Delta T_{EC}$ for both PLZT compositions are shown in Fig. 3 at a few selected amplitudes of the electric field. To preserve the sample and not to exceed the breakdown field, the electric field strength was limited to a maximum of 60-80 kV/cm.

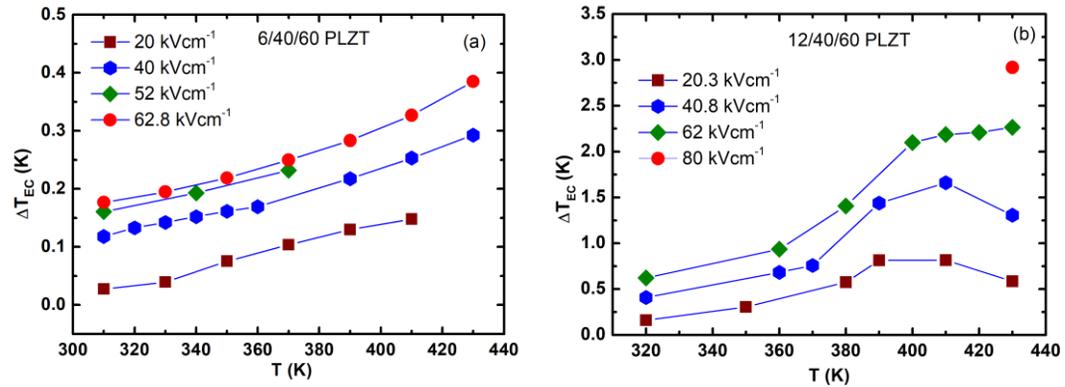

FIG. 3. The ECE temperature change $\Delta T_{EC}$ in 6/40/60 PLZT bulk ceramics (a) and 12/40/60 PLZT ceramics (b) as a function of temperature for several amplitudes of the electric field.

The temperature dependence of the ECE specific entropy change $\Delta s$ for both PLZT compositions are shown in Fig. 4. Specific entropy change $\Delta s$ is calculated by scaling the total measured ECE entropy change $\Delta s = \Delta S/m = c_{ph}\Delta T/T$ by the mass of the sample $m$ [46,47]. Here, phonon specific heat values of $c_{ph}$ = 330 and 350 J/kgK were taken into account for 6/40/60 and 12/40/60 PLZT ceramics, respectively. Note that in the temperature range of interest, the $c_{ph}$ is a very week function of temperature [47].

Results presented in Figs. 3 and 4 demonstrate that the ECE is positive in the measured temperature range, even for lower electric fields as expected for the ferroelectric phase. The temperature dependence and the magnitude of the ECE response determined in 6/40/60 PLZT bulk ceramics (Figs. 3(a) and 4(a)) are qualitatively similar to those observed in 12/40/60 PLZT ceramics well below the transition temperature (Figs. 3(b) and 4(b)), i.e., gradually increasing dependence with increasing temperature. The maximum value of $\Delta T_{EC}$ observed in 6/40/60

PLZT bulk ceramics was 0.39 K at 430 K and 62.8 kV/cm. This value is similar to values found in 12/40/60 PLZT ceramics near the room temperature, about 100 K below the FE transition. In contrast, in 12/40/60 PLZT ceramics, temperature dependence demonstrates that the ECE response at the FE transition is enhanced, with the maximum value of $\Delta T_{EC}$ = 2.92 K at 430 K and 80 kV/cm. It should be noted that by increasing the applied electric field, the FE transition is shifted toward higher temperatures, as observed in other perovskite ferroelectrics [30].

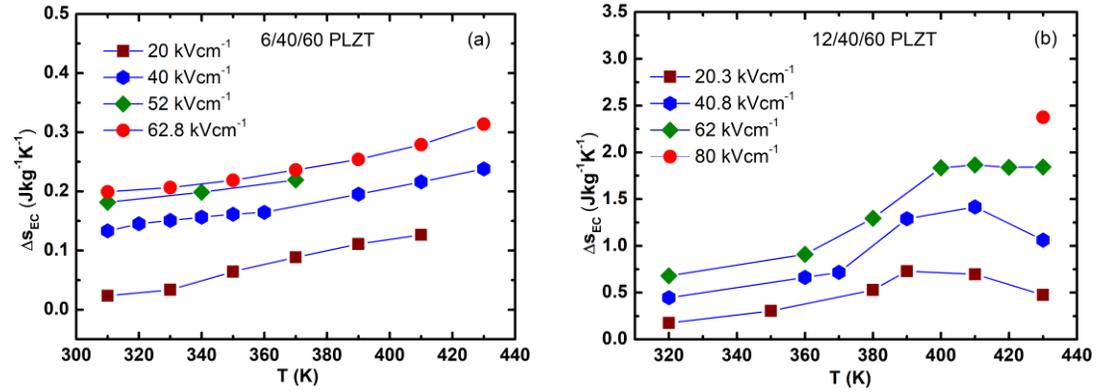

FIG. 4. The ECE specific entropy change $\Delta s$ in 6/40/60 PLZT bulk ceramics (a) and 12/40/60 PLZT ceramics (b) as a function of temperature for several amplitudes of the electric field.

The above maximum value of $\Delta T_{EC}$ measured in 12/40/60 PLZT ceramics exceeds previously published values in all La-modified x/65/35 PLZT compositions [1,8,30] that are more relaxor type with rather small latent heat [32]. $\Delta T_{EC}$ is close to the 3.45 K observed in PMN-10PT, but at a much higher electric field of 140 kV/cm [36]. We argue that such ECE enhancement for relatively modest electric fields is most likely due to the latent heat of the first order P-FE transition in 12/40/60 PLZT ceramics [45,47]. Much sharper rise of the $\Delta T_{EC}$ near the FE transition than in 8/65/35 PLZT ceramics is signaling that the $\Delta T_{EC}$ is enhanced by the latent heat up to $L/c_{ph} \approx 0.7$ K. Here, the latent heat ($L$) of the FE transition in 12/40/60 PLZT ceramics, is estimated from calorimetric measurements to $L \cong 250$ J/kg [47]. Typical latent heat values of 60 J/kg to 400 J/kg, such as observed in Lead Magnesium Niobate (PMN) and Barium Titanate (BTO), respectively, can enhance the ECE for 0.2 to 1 K [15,29]. The value of the electrocaloric responsivity determined at lower electric fields of 20 kV/cm $\Delta T/\Delta E$=0.41x10$^{-6}$ Km/V at 410 K is comparable to those observed in other perovskite ferroelectrics [1,9,12,15,17,30,48-50]. The value of the $\Delta T/\Delta E$=0.37x10$^{-6}$ Km/V at a high electric field of 80 kV/cm at 430 K remains high and is one of the highest at such a strong field in perovskite ferroelectrics.

## IV. CONCLUSION

In conclusion, we have explored by direct measurements the temperature and electric field dependence of the electrocaloric response in bulk 6/40/60 and 12/40/60 PLZT ceramics. The electrocaloric effect was investigated in the temperature range near the room temperature, i.e., from 310 K up to 430 K. Relatively small value of ECE response not exceeding 0.4 K at 62.8 kV/cm were found in 6/40/60 PLZT sample. In contrast, in the 12/40/60 PLZT sample, one of the highest reported $\Delta T_{EC}$ values of 2.92 K at 430 K and 80 kV/cm was observed due to the paraelectric-ferroelectric phase transition latent heat enhancement. The results demonstrated the importance of the proximity of the first-order paraelectric to ferroelectric phase transition for electrocaloric response enhancement and that La-modified x/40/60 PLZT compositions exhibit significant electrocaloric response.

The data that supports the findings of this study are available within this article.

## ACKNOWLEDGMENTS

This work was supported by the project 778072 — ENGIMA — H2020-MSCA-RISE-2017, Slovenian Research Agency grant J1-9147, and program P1-0125. The authors would like to acknowledge critical help in sample preparation by Jena Cilenšek and Silvo Drnovšek.

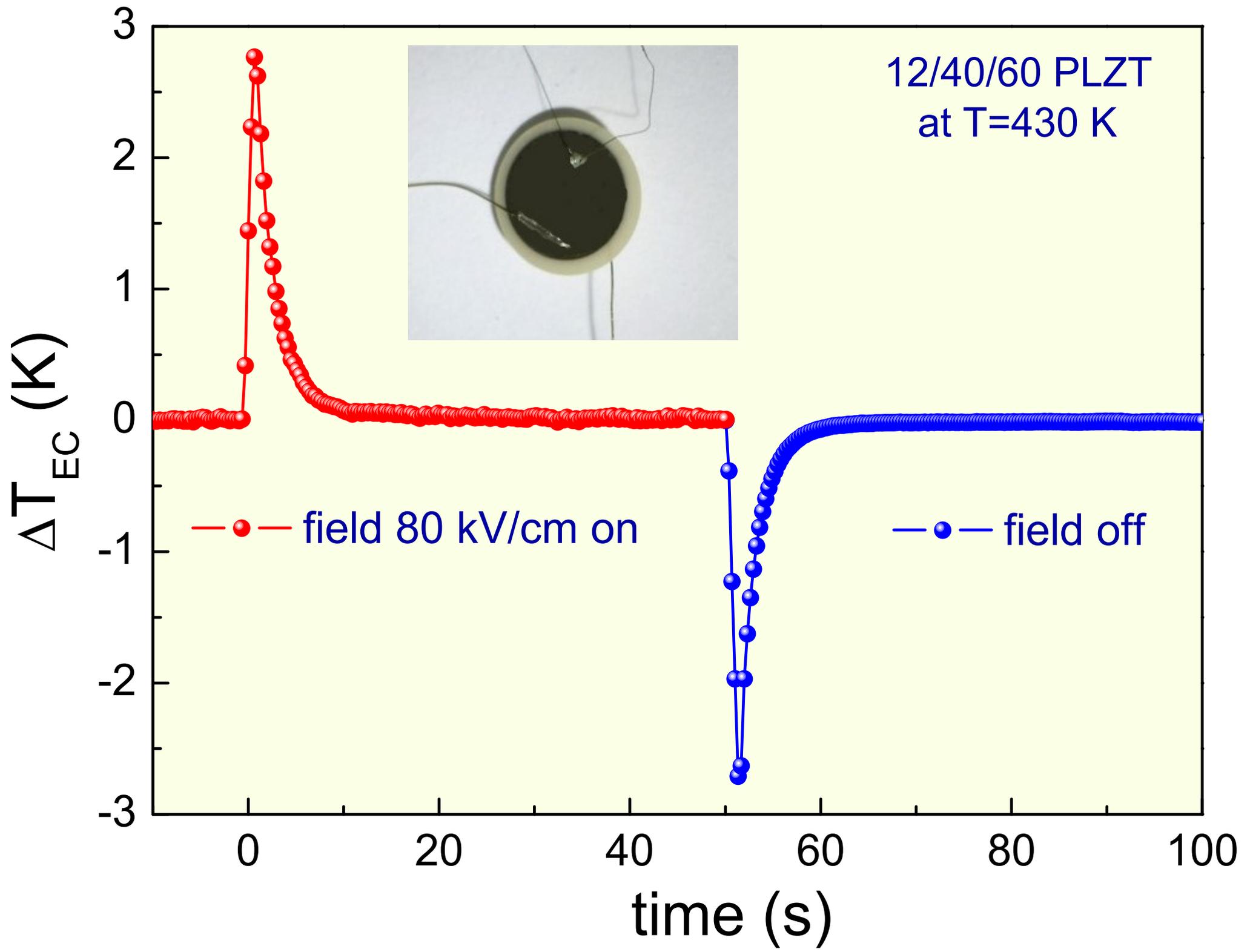

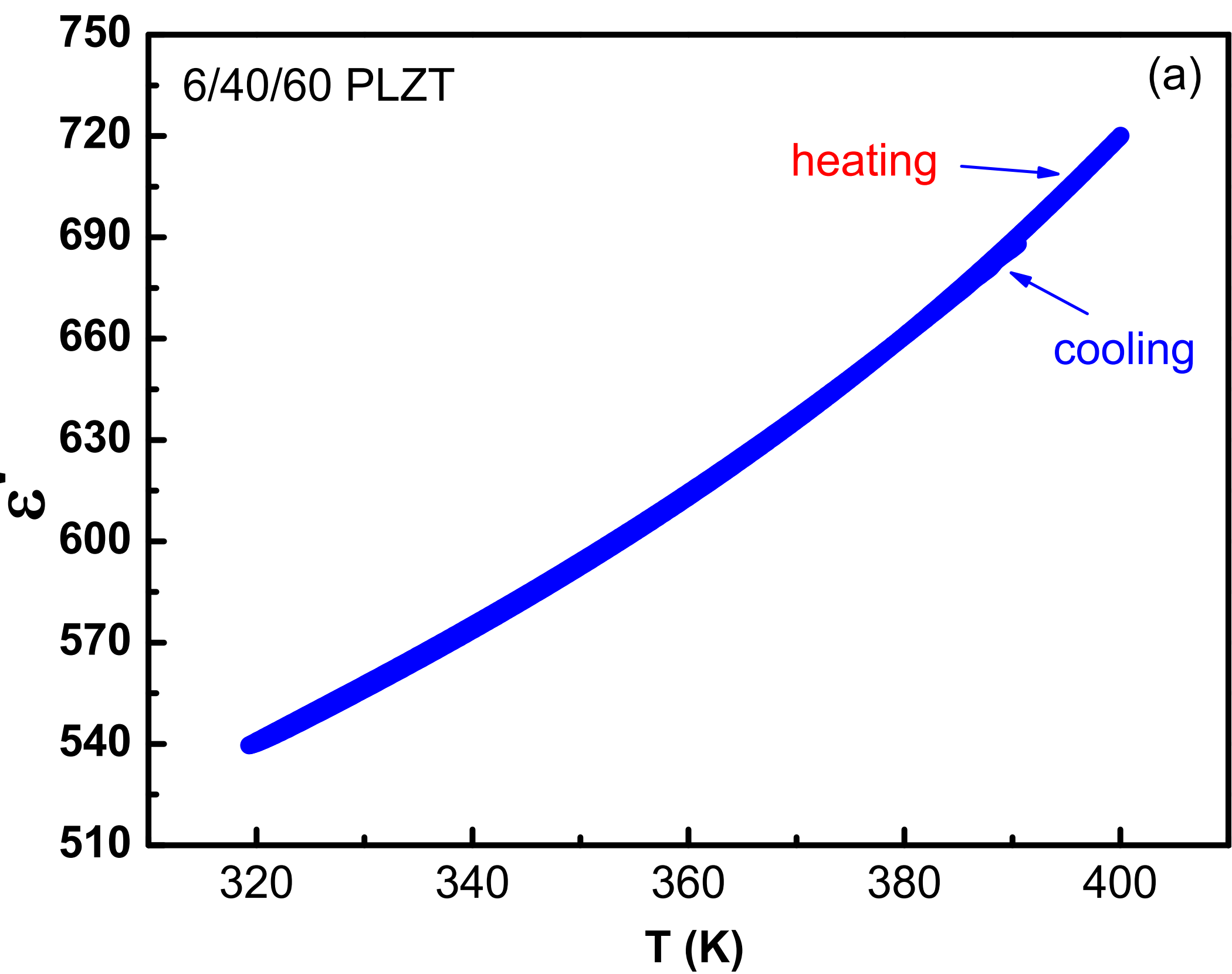

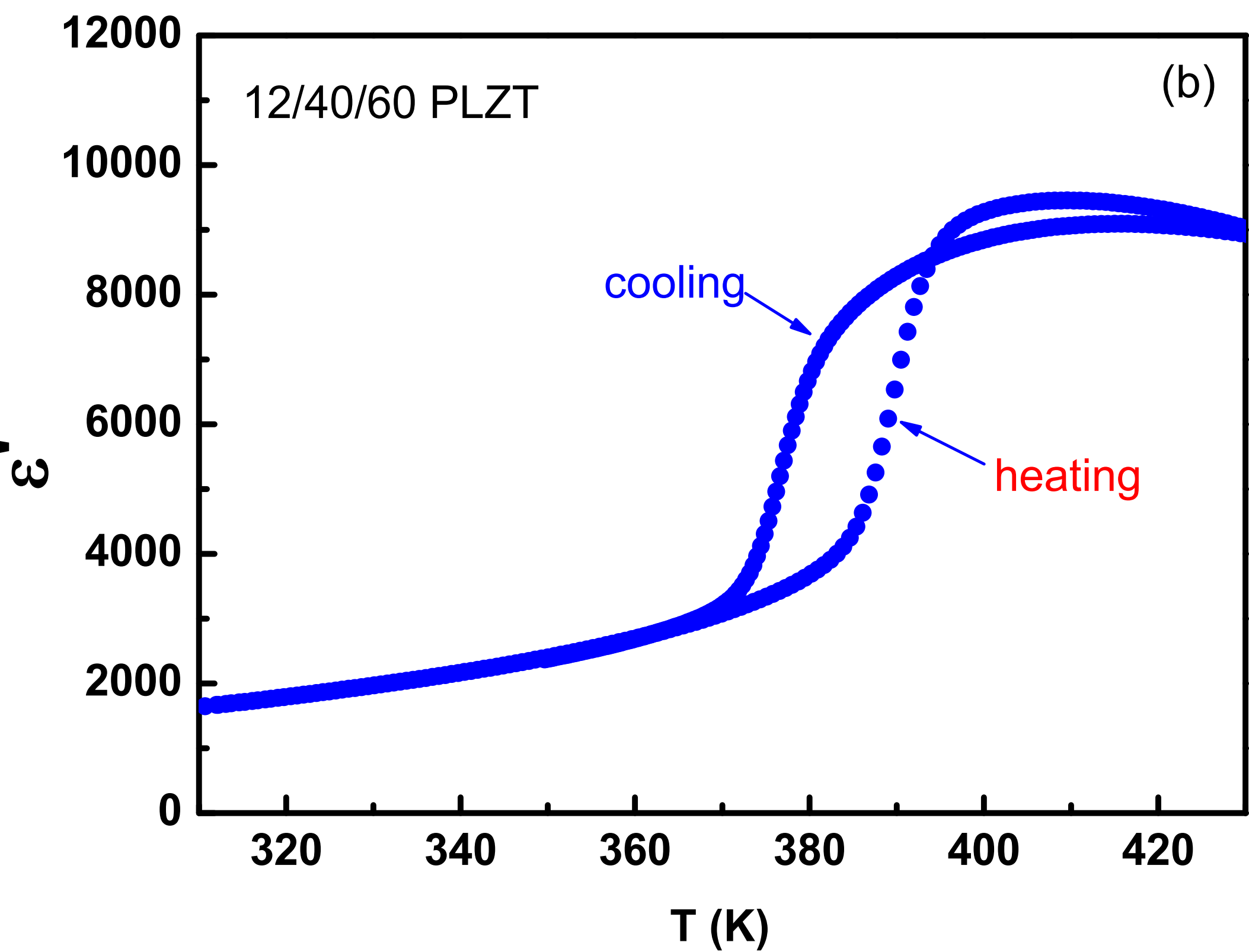

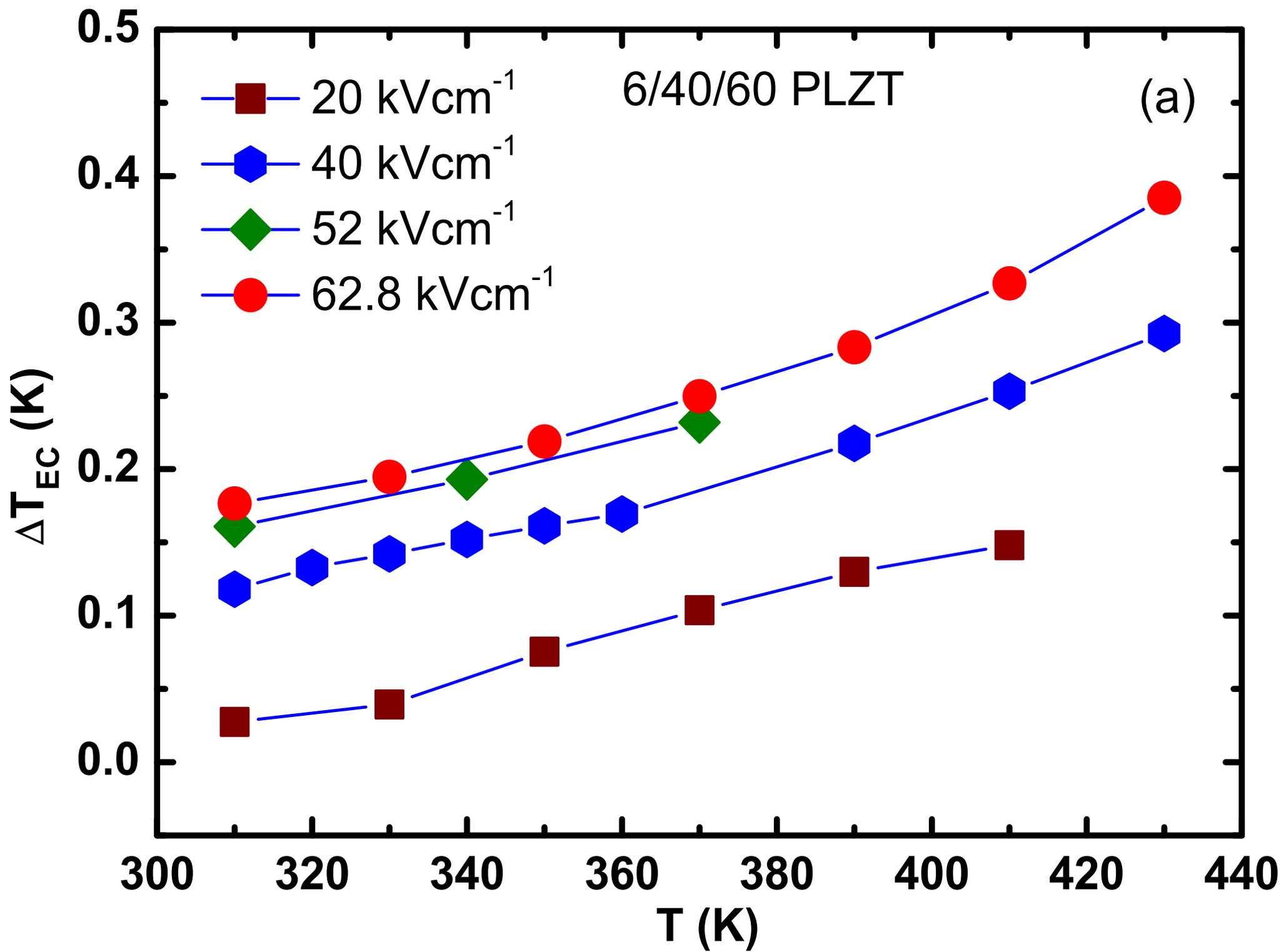

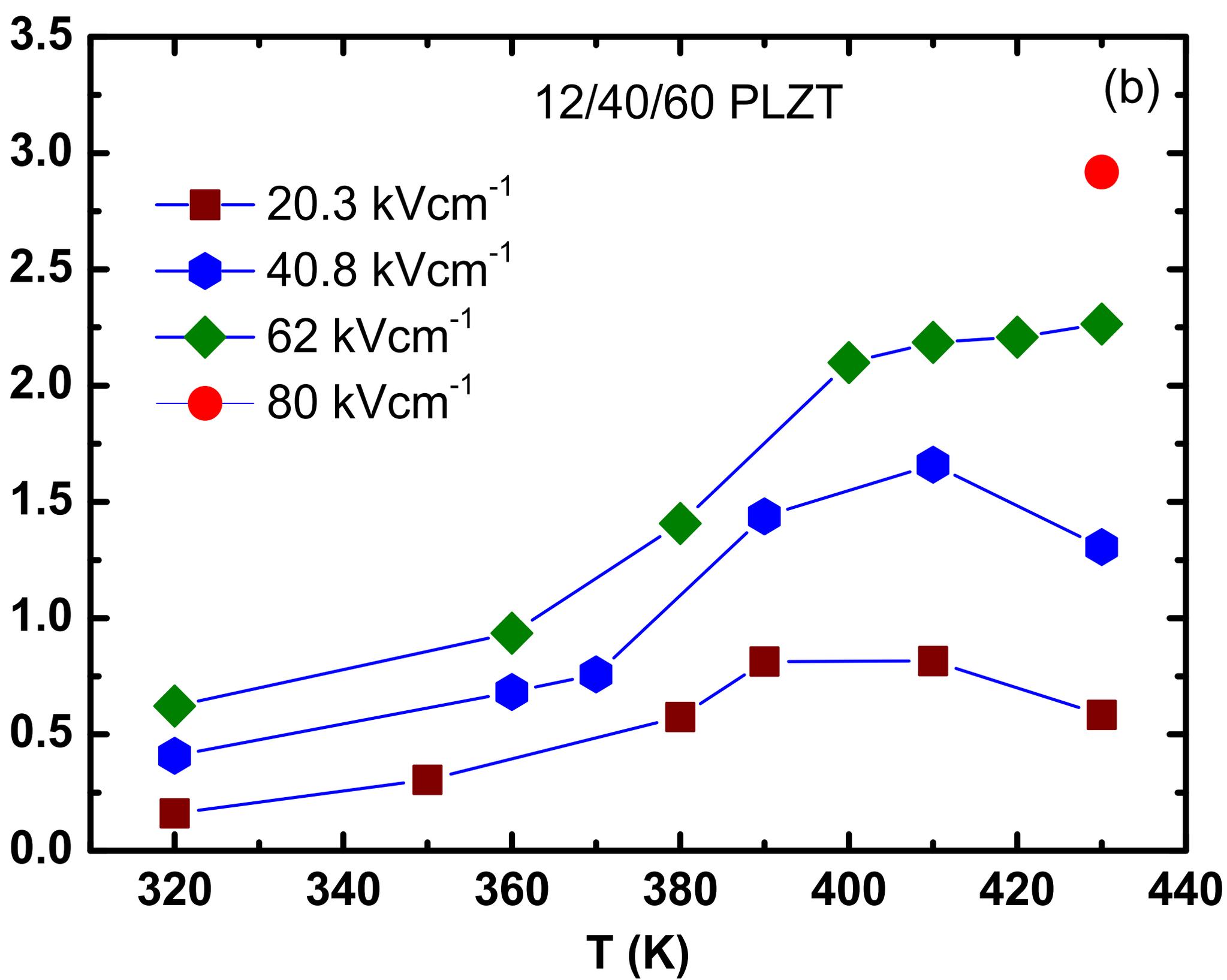

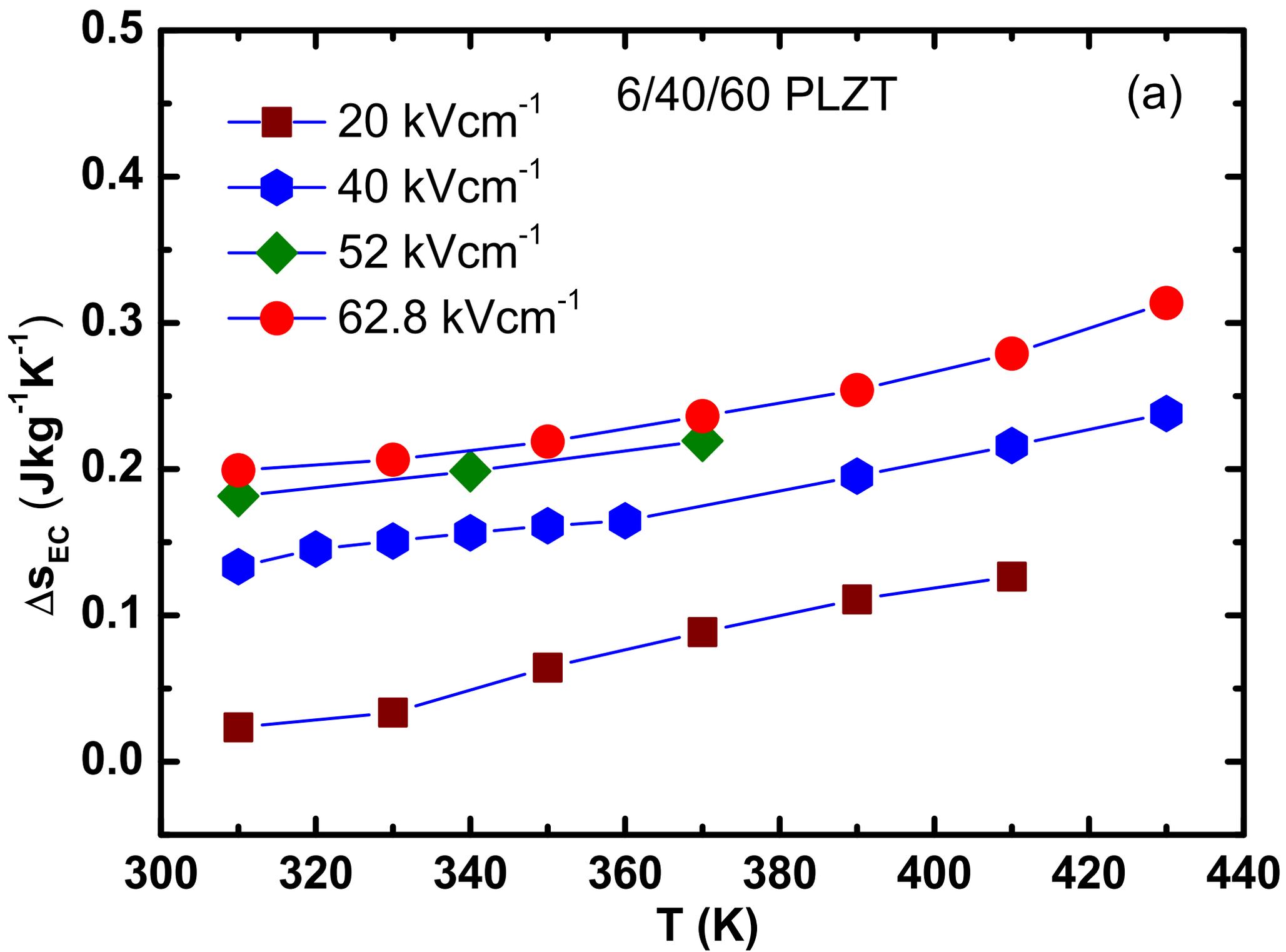

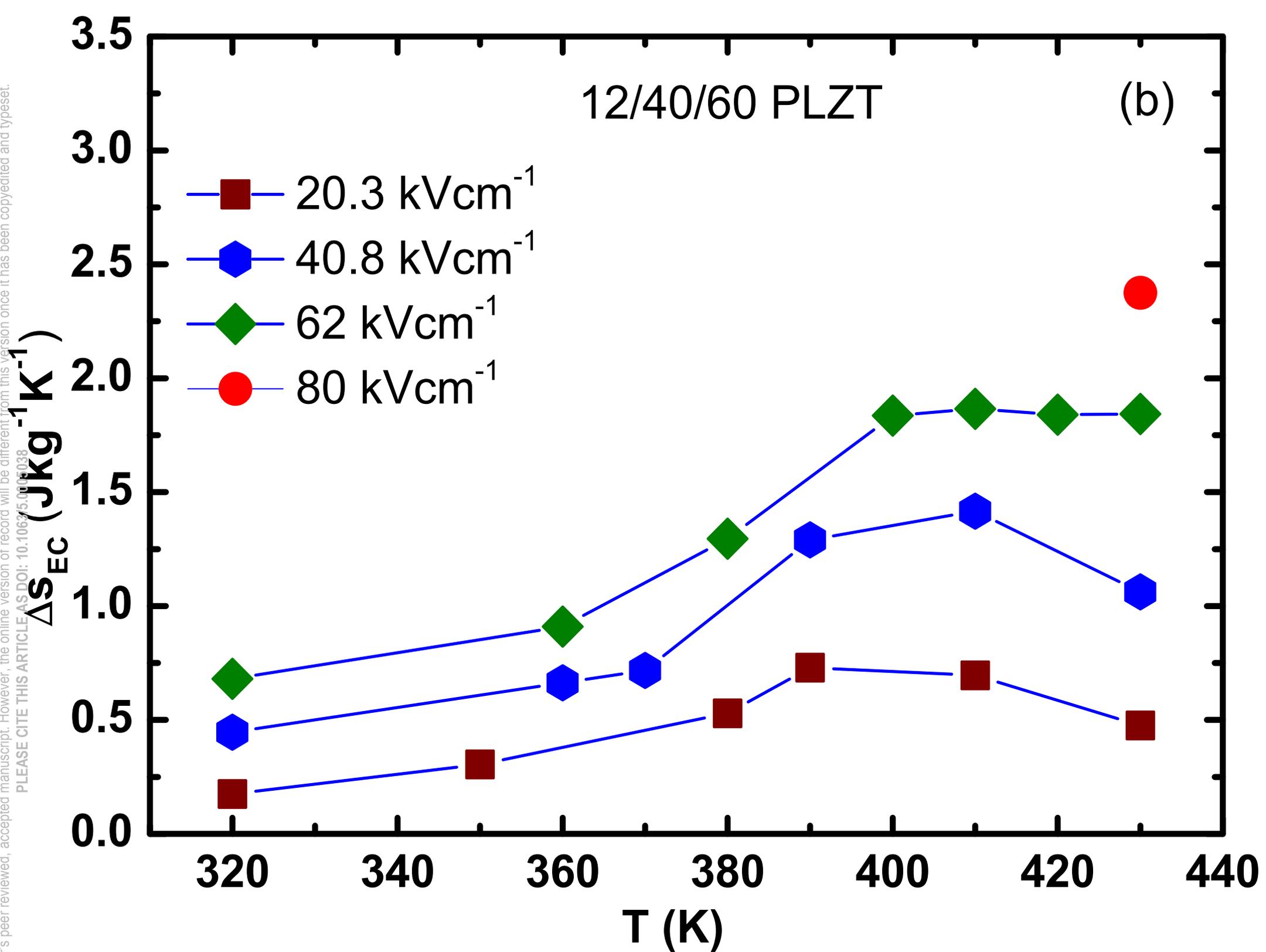